\documentstyle[12pt]{article}
\newcommand{\nn}{\nonumber}
\def\bbox#1{{\bf #1}}
\def\text#1{{\rm #1}}
\newcommand{\wt}{\widetilde}
\newcommand{\D}{{\underline{\Delta}}}

\def\Journal#1#2#3#4{{#1} {#2} (#4) #3}
\def\Book#1#2#3#4{{#1}, #2 (#3, #4)}
\def\Proc#1#2#3#4#5#6{{#1}, in: #3 (#5, #6) p.~#4}

\def\PLA{{Phys. Lett.}  A}
\def\JMP{J. Math. Phys.}
\def\LOMI{Zap. Nauchn. Semin. LOMI}
\def\FAP{Funkt. Anal. Ego Pril.}
\def\PHD{Physica D}
\def\LMP{Lett. Math. Phys.}

\def\be{\begin{equation}}
\def\ee{\end{equation}}
\def\bea{\begin{eqnarray}}
\def\eea{\end{eqnarray}}

\begin{document}

\title{Generalized KP hierarchy:\\
M\"obius Symmetry, Symmetry Constraints
\\and Calogero-Moser System}

\author{L.V. Bogdanov\thanks{
L.D. Landau ITP, Kosygin str. 2,
Moscow 117940, Russia; e-mail Leonid@landau.ac.ru}
\hspace{0.1em} and B.G. Konopelchenko
\thanks{
Dipartimento di Fisica dell' Universit\`a
and Sezione INFN, 73100 Lecce, Italy}}


\maketitle
\abstract{Analytic-bilinear approach is used to study
continuous and discrete
non-isospectral symmetries of the generalized KP
hierarchy.
It is shown that M\"obius symmetry transformation
for the singular manifold equation leads to continuous or discrete
non-isospectral symmetry of the basic (scalar or multicomponent KP)
hierarchy connected with binary B\"acklund transformation. A more
general class of multicomponent M\"obius-type symmetries is studied.
It is demonstrated that symmetry constraints of KP hierarchy
defined using multicomponent M\"obius-type symmetries give rise
to Calogero-Moser system.}

\section{Introduction}
It is a pleasure for us to dedicate this paper to Prof.
V.E. Zakharov 60th birthday. The technique used in this work
(analytic-bilinear approach, see \cite{AB1}, \cite{AB2})
takes its origin in the ideas of the $\bar{\partial}$-dressing
method developed by V.E. Zakharov and S.V. Manakov \cite{dbar}.
It is interesting to note that the first paper formulating
basic ideas of analytic-bilinear approach \cite{NLS} was
published in the volume dedicated to 55th birthday of
Prof. Zakharov.

Analytic-bilinear approach formalizes important features of the
$\bar{\partial}$-dressing method connected with construction
of integrable equations, leaving aside some details of the scheme
of generating special classes of solutions. In this form the method
becomes close to Grassmannian approach \cite{Sato}, \cite{Wilson},
thus filling the gap between $\bar{\partial}$-dressing method
and more abstract Grassmannian approach,
preserving at the same time some useful structures characteristic
of the original  $\bar{\partial}$-dressing method.

In this work we use analytic-bilinear approach to study
continuous and discrete
non-isospectral symmetries of the generalized KP
hierarchy. We demonstrate that M\"obius symmetry on the
level of KP singular manifold equations (KPSM) hierarchy,
binary B\"acklund transformations of KP hierarchy
and solitonic transformations of the $\tau$-functions
through the Date-Jimbo-Kashiwara-Miwa vertex operator
are different manifestations of the same discrete symmetry.
Considering continuous non-isospectral symmetries, we show that
Calogero-Moser system can be obtained through the symmetry constraint
of KP hierarchy.

\section{Generalized KP Hierarchy}
First we give a sketch of the picture of generalized KP
hierarchy in frame of analytic-bilinear approach;
for details we refer to \cite{AB1}, \cite{AB2}.

The formal starting point is Hirota bilinear identity
for Cauchy-Baker-Akhiezer function,
\bea
\oint \chi(\nu,\mu;g_1)g_1(\nu)g_2^{-1}(\nu)
\chi(\lambda,\nu;g_2)d\nu=0\,\quad \lambda\,,\mu\in D.
\label{HIROTA0}
\eea
Here $\chi(\lambda,\mu;g)$ (the Cauchy kernel)
is a function of two complex variables
$\lambda ,\mu\in \bar D$, where $D$ is a unit disc,
and a functional of the loop group
element $g\in\Gamma^+$, i.e., of a
complex-valued function analytic and having
no zeros in $\bbox{C}\setminus D$, equal to 1 at infinity;
the integration goes over the
unit circle . By definition, the function $\chi(\lambda,\mu)$
possesses the following analytical properties: as $\lambda
\rightarrow\mu$, $\chi\rightarrow (\lambda-\mu)^{-1}$ and
$\chi(\lambda,\mu)$ is an analytic function of
two variables $\lambda,\mu \in \bar D$ for $\lambda\neq\mu$.
The function $\chi(\lambda,\mu;g)$ is a solution to
(\ref{HIROTA0}) if it possesses specified analytic properties
and satisfies (\ref{HIROTA0}) for all $\lambda,\mu \in D$ and
some class of loops $g\in\Gamma^+$.

In  another form, more similar to standard Hirota bilinear identity,
the identity (\ref{HIROTA0}) can be written as
\be
\oint \psi(\lambda,\nu;g_2)
\psi(\nu,\mu;g_1)d\lambda=0 ,
\ee
where
$$
\psi(\lambda,\mu,g)=g(\lambda)\chi(\lambda,\mu,g)g^{-1}(\mu).
$$
We call the function $\psi(\lambda,\mu;g)$ a Cauchy-Baker-Akhiezer
function.

Hirota bilinear identity (\ref{HIROTA0})
incorporates the standard Hirota bilinear identity
for the Baker-Akhiezer (BA) and dual (adjoint) Baker-Akhiezer function
of the KP hierarchy.
Indeed, let us introduce
these functions by the formulae
\bea
\psi(\lambda;g)=g(\lambda)\chi(\lambda;0),
\nn\\
\wt\psi(\mu;g)=g^{-1}(\mu)\chi(0;\mu).
\nn
\eea
Then for the Baker-Akhiezer function
$\psi(\lambda;g)$ and the dual Baker-Akhiezer
function
$\wt\psi(\mu;g)$,
taking the identity (\ref{HIROTA0})
at $\lambda=\mu=0$, we get the usual form of the Hirota bilinear identity
\be
\oint \wt\psi(\nu;g_2)
\psi(\nu,;g_1)d\nu=0.
\ee
The only minor difference from the standard setting here is that
we define the BA and dual BA function in the neighborhood of zero,
not in the neighborhood of infinity.

There are three different types of integrable
discrete equations implied by identity (\ref{HIROTA0}),
that, in the continuous limit, lead
to the KP hierarchy in the usual form (in terms of potentials),
to the modified KP hierarchy and to the hierarchy of the singular manifold
equations. They arise for different types of functions
connected with the Cauchy-Baker-Akhiezer function
satisfying Hirota bilinear
identity (see the derivation in \cite{AB1}, \cite{AB2}).

On the first level,
we have the equations for the diagonal of the regularized
Cauchy kernel taken at zero (the potential)
$$
u(g)=\left(\chi(\lambda,\mu;g)-
(\lambda-\mu)^{-1}\right)_{\lambda=0\,,\mu=0}\,,
$$
on the second level, the equations for the
Baker-Akhiezer and dual Baker-Akhiezer type wave
functions (the modified equations)
$$
\Psi(g)=\int\psi(\lambda,g) \rho(\lambda)d\lambda\,,
$$
$$
\widetilde \Psi(g)=
\int\widetilde\rho(\mu)\widetilde\psi(\mu,g) d\mu\,,
$$
and on the third level -- the equations for the
Cauchy-Baker-Akhiezer type wave function
\bea
\Phi(g)=\oint\!\!\oint (\psi(\lambda,\mu;g))
\rho(\lambda)
\wt\rho(\mu) d\lambda\,d\mu,
\nn
\eea
where $\rho(\lambda)$,
$\wt\rho(\mu)$ are some arbitrary weight functions.
The equations of all three levels
possess an infinite number of commuting symmetries
and form in some sense a hierarchy of integrable
discrete equations represented in the form of the general equation
labeled by three continuous parameters (the lattice parameters).

To present discrete equations forming three levels
of generalized KP hierarchy, we introduce difference
and shift operators
$$
T_a f(g)=f(g\times g_a^{-1}),
$$
$$
\Delta_a =T_a-1,
$$
$$
\widetilde T_a f(g)=T_a^{-1} f(g)=f(g\times g_a),
$$
$$
\widetilde\Delta_a =1-\widetilde T_a,
$$
where elementary rational loop $g_a(\nu)$
is defined as
$$
g_a(\nu):=
{\frac{\nu-a
}{\nu}}
$$
We will use shift and difference operators with different values of
lattice parameter $a=a_i$ denoting
$$
T_i=T_{a_i}.
$$

The first level of the generalized KP hierarchy is
formed by equations for the potential
$u$
\bea
\sum_{(ijk)} \epsilon_{ijk}T_k \left({\Delta_i\over a_i} u
- uT_iu\right)=0,
\label{KPa0}
\eea
where $i\neq j\neq k\neq i$; $i,j,k\in\{1,2,3\}$,
summation goes over different permutations of indices.
Hirota bilinear identity (\ref{HIROTA0}) also implies
linear equations
\begin{eqnarray}
&&
\left({\Delta_i\over a_i}-{\Delta_j\over a_j} \right) \widetilde\Psi(g)
=
\left((T_i-T_j)u(g)\right)\widetilde\Psi(g)
\label{compa}
\,
\end{eqnarray}
and
\bea
&&
\left({\widetilde\Delta_i\over a_i} -
{\widetilde\Delta_j\over a_j} \right)\Psi(g)
=
\left((\widetilde T_i-\widetilde T_j)u(g)\right)
\Psi(g)
\label{compb}
\,,
\end{eqnarray}
that generate equation (\ref{KPa0})
as compatibility condition

The second
level of the generalized hierarchy is formed
by equations for the wave functions
of linear operators producing equations of the first
level as compatibility conditions (the modified equations).
It splits into two
parts: equations for the dual wave function $\wt\Psi$
\bea
\sum_{(ijk)}\epsilon_{ijk}a_ja_kT_k\left(\widetilde \Psi^{-1}
({T_i}\widetilde \Psi)
\right)=0,
\label{KPwavea1}
\eea
and equations for the wave functions $\Psi$
\bea
\sum_{(ijk)}\epsilon_{ijk}a_ja_k\wt T_k\left(\Psi^{-1}
({\wt T_i}\Psi)
\right)=0.
\label{KPwaveb1}
\eea

The third level represents equations for the
wave functions $\Phi$ of linear operators of the second level
(the singular manifold type equations)
\bea
(T_j\Delta_i \Phi)(T_k\Delta_j \Phi)(T_i\Delta_k \Phi)=
(T_j\Delta_k \Phi)(T_k\Delta_i \Phi)(T_i\Delta_j \Phi).
\label{singmana}
\eea
One could expect
this chain to continue, but the wave functions
of linear operators of the third level coincide
with the wave functions of linear operators of the first level,
and so the chain closes.

Equations of the second and third levels of the hierarchy and linear
problems for them can be derived from the set of simple equations
that directly follows from identity (\ref{HIROTA0}), namely
\bea
{\Delta_i\over a_i}\Phi=\wt\Psi T_i\Psi
\label{Phi1}
\eea
and, equivalently,
\bea
{\wt\Delta_i\over a_i}\Phi=\Psi \wt T_i\wt \Psi.
\label{Phi2}
\eea

To reproduce second and third levels of the generalized KP hierarchy
from the first,
taking as a basic object equation (\ref{KPa0})
and without reference to bilinear identity,
it is enough to
notice that
linear equations
(\ref{compa}) and (\ref{compb}) imply that there exists a
function $\Phi(\bbox{x})$ satisfying equations (\ref{Phi1}),
(\ref{Phi2}).
Indeed, using equations (\ref{compa}), (\ref{compb})
it is easy to check that
cross-differences for the set of equations (\ref{Phi1}), (\ref{Phi2})
are equal, and the function $\Phi$ is well-defined on the
lattice (and through some limit also as a function of continuous
variables).

Connections between three different levels of the hierarchy
of discrete equations may be described in terms of
Miura maps and Combescure symmetry transformations,
which are in some sense complementary.

\section{From Discrete Equations to the Continuous Hierarchy}

The loop $g\in\Gamma^+$ can be parametrized by the infinite
set of complex variables $x_i$, $1\leq i\leq\infty$,
$$
g(\lambda)=\exp(\sum_{i=1}^{\infty} x_i\lambda^{-i}),
$$
and then the functionals of $g$ may be considered as
functions of the infinite set of variables
$$
\bbox{x}=\{x_1,x_2,\dots,x_n,\dots\}.
$$
The transformation operators $T_i$
now look like
$T_i:{\bf x}\rightarrow{\bf x}+[a_i]$,
where
$[a]_n={1\over n}a^n$. To compactify the notations, we will also
use scaled difference operators
$$\D_i=a_i^{-1}\Delta_i,\quad
\wt\D_i=a_i^{-1}\wt\Delta_i.$$
The transformation operators $T_a$ can be represented in terms
of differential operators in the form
$
T_a=\sum_{n=1}^{\infty}a^n p_n(\widetilde\partial)
$,
$
\widetilde T_a=\sum_{n=1}^{\infty}a^n p_n(-\widetilde {\partial})
$,
where
$
{\widetilde\partial} =\left({\frac{\partial}{
\partial x_1}},{\frac{1}{2}}{\frac{\partial}{\partial x_2}}, \dots,
{\frac{1
}{n}}{\frac{\partial}{\partial x_n}},\dots\right)
$,
and $p_i$ are the Schur polynomials generated by the
relation
$
\exp\left(\sum_{n=1}^{\infty} \lambda^n x_n\right)=
\sum_{n=0}^{\infty}p_i(\bbox{x})\lambda^n
$.
For the first three continuous variables we will use the notations
$x=x_1$, $y=x_2$, $t=x_3$.

To demonstrate that the discrete form of KP hierarchy (\ref{KPa0})
written in terms of continuous variables as
\begin{equation}
\sum_{(ijk)} \epsilon_{ijk}T_k \left({\Delta_i\over a_i} u(\bbox{x})
- u(\bbox{x})T_iu(\bbox{x})\right)=0
\label{KPa1}
\end{equation}
generates equations of KP hierarchy in the standard form, we consider
expansion of this equation into powers of parameters $a_i\,,a_j\,,a_k$.
The zeroth order of expansion
of equation (\ref{KPa1}) into powers of the
parameter $a_i$ gives the equation
containing two discrete transformations and one partial derivative
\bea
&&
T_k \left(\partial_x u
- uu\right)-
\left({\D_k} u
- uT_ku\right)
\nn\\&&
\qquad+\left({\D_j} u
- uT_ju\right)-
T_j \left(\partial_x u
- u u\right)
\nn\\&&
\qquad\qquad+
T_j \left({\D_k} u
- uT_ku\right)-
T_k \left({\D_j} u
- uT_ju\right)=0.
\label{KPdiscrete1}
\eea
The first order of expansion of equation (\ref{KPdiscrete1})
into the powers of the parameter $a_j$
represents  an equation containing
partial derivatives over two continuous variables and one discrete
transformation,
\bea
&&
\left(\mbox{${1\over2}$}\left(\partial_y+\partial_x^2\right) u
- u\partial_x u\right)-
\partial_x \left(\partial_x u
- u u\right)
\nn\\&&
\qquad+
\partial_x \left({\D_k} u
- uT_ku\right)-
T_k \left(\mbox{${1\over2}$}\left(\partial_y+\partial_x^2\right) u
- u\partial_x u\right)=0.
\label{KPdiscrete2}
\eea
The final step is to take the first nontrivial
order of the expansion into the powers of $a_k$
(the second order) to get
the potential form of the KP equation
\begin{equation}
\partial_x\left(u_t-\mbox{${1\over 4}$}u_{xxx}+
\mbox{${3\over2}$}(u_x)^2\right)=\mbox{${3\over4}$}
u_{yy}\,,
\label{KP00}
\end{equation}
which reduces to standard KP equation for
the function $v=-2\partial_x u$.

The higher orders of expansion of equation
(\ref{KPdiscrete2}) will give us the higher equations of the
KP hierarchy. This sequence of equations should be used recursively
to get equations containing only partial derivatives
over the highest order time, $\partial_x$ and $\partial_y$.

The interpretation of the chain of equations we have derived
depends on the
choice of the basic equation (i.e., in some
sense on the point of reference).

A standard way is to take continuous equation
(\ref{KP00}) (or, rather, the KP hierarchy
in the form of PDEs)
as a basic system. Then the interpretation
of the other objects is:
1) equation (\ref{KPdiscrete2}) defines a B\"acklund
transformation for equation (\ref{KPa1}),
2) equation (\ref{KPdiscrete1}) is a superposition
principle for two B\"acklund transformations,
3) discrete  equation (\ref{KPa1})
provides an {\em algebraic} superposition principle
for three B\"acklund transformations.

On the other hand, the discrete equation
(\ref{KPa1}) (in other words, the discrete form of the
KP hierarchy)
may be treated as a basic system as well.
Then formula (\ref{KPdiscrete1}) is a lowest order continuous
symmetry for this system, equation (\ref{KPdiscrete2})
is a superposition principle for two continuous symmetries,
and equation (\ref{KP00}) is a superposition
principle for three continuous symmetries of different orders.

Linear problems (\ref{compa}), (\ref{compb})
generating the discrete form of the KP
hierarchy (\ref{KPa1}) as compatibility conditions
in terms of continuous variables look like
\begin{eqnarray}
&&
({\D_i}-{\D_j}) \widetilde\Psi(\bbox{x})
=
((T_i-T_j)u(\bbox{x}))\widetilde\Psi(\bbox{x})
\label{KPbasea13}
\,,
\\
&&
({\widetilde\D_i} -
{\widetilde\D_j})\Psi(\bbox{x})
=
((\widetilde T_i-\widetilde T_j)u(\bbox{x}))
\Psi(\bbox{x})
\label{KPbaseb13}
\,.
\end{eqnarray}
Both the set of equations (\ref{KPbasea13}) and the dual set
(\ref{KPbaseb13}) imply the same equation (\ref{KPa1}). Expansion
of these linear equations into powers of parameters gives standard
linear problems for KP hierarchy.

We will not use equations of the
second level of generalized KP hierarchy in the present work.
Equation for $\Phi$ (\ref{singmana})
(discrete form of KP singular manifold equation
hierarchy) in terms of continuous variables looks like
\bea
&&
(T_j\Delta_i \Phi(\bbox{x}))(T_k\Delta_j \Phi(\bbox{x}))
(T_i\Delta_k \Phi(\bbox{x}))
\nn\\
&&
\qquad=
(T_j\Delta_k \Phi(\bbox{x}))(T_k\Delta_i \Phi(\bbox{x}))
(T_i\Delta_j \Phi(\bbox{x})),
\label{KPSM}
\eea
and, performing expansion into powers of parameters, we get
the chain of equations connecting discrete and continuous case,
\begin{equation}
(T_j \Phi_x)(T_k\Delta_j\Phi)\Delta_k\Phi=
(T_k\Phi_x)(T_j\Delta_k\Phi)\Delta_j\Phi,
\label{SM2-1}
\end{equation}
\begin{equation}
{\partial\over \partial x}
\ln \left({1\over \Phi_x}
{\Delta \Phi\over a}\right)=
{1\over 2}\Delta\left({\Phi_y+\Phi_{xx}\over
\Phi_x}\right)
\label{SM1-2},
\end{equation}
\begin{eqnarray}
&&\Phi_t=\mbox{${1\over4}$}\Phi_{xxx}+\mbox{${3\over8}$} {\frac{
\Phi_y^2-\Phi_{xx}^2}{\Phi_x}}+ \mbox{${3\over4}$}\Phi_x W_y ,
\quad W_x={
\frac{\Phi_y}{\Phi_x}}\,.
\label{singman1}
\end{eqnarray}
The last equation first arose in Painleve analysis of the KP
equation as a
singular manifold equation \cite{Weiss}.

The interpretation of this chain of equations is similar to
the interpretation given
for equation (\ref{KPa1}).

Thus the integrable discrete equations
written in terms of elementary
rational loops encode the continuous hierarchy,
the B\"acklund transformations and different
types of superposition principles for them,
and the discrete linear equations generate a
hierarchy of linear problems,
Darboux transformations and superposition principles for them.

\section{Discrete and Continuous Non-Iso\-spec\-tral Symmetries}
The dynamics defined by Hirota bilinear
identity (\ref{HIROTA0}) is connected with operator of
multiplication by loop group element $g\in \Gamma^+$;
this dynamics can be interpreted in terms of commuting flows
corresponding to infinite number of `times'
$x_n$. A general idea of introduction additional
(in general, non-commutative) symmetries is to consider
more general operators $\hat R$ on the unit circle.
Let us introduce symmetric bilinear form
$$
(f|g)=\oint f(\nu)g(\nu)d(\nu).
$$
In terms of this form identity (\ref{HIROTA0})
looks like
\bea
(\chi(\dots,\mu;g_1)g_1(\dots)|
g_2^{-1}(\dots)\chi(\lambda,\dots;g_2))=0\,\quad \lambda\,,\mu\in D,
\label{HIROTA01}
\eea
or, for Cauchy-Baker-Akhiezer function $\psi(\lambda,\mu;g)$,
\bea
(\psi(\dots,\mu;g_1)|
\psi(\lambda,\dots;g_2))=0\,\quad \lambda\,,\mu\in D,
\label{HIROTA02}
\eea
where by dots we denote the argument which is involved into
integration.
Let some CBA function $\psi(\lambda,\mu;g)$ satisfying Hirota
bilinear identity be given.
We define symmetry transformation connected with
arbitrary invertible
linear operator $\hat R$ in the space of functions on the
unit circle by the equations
\bea
(\wt\psi(\dots,\mu;g_1)|\hat R|
\psi(\lambda,\dots;g_2))=0,
\nn\\
(\psi(\dots,\mu;g_1)|\hat R^{-1}|
\wt\psi(\lambda,\dots;g_2))=0.
\nn
\eea
It is possible to show that if both these equations
for the transformed CBA function $\wt\psi(\lambda,\mu;g)$
are solvable, then the solution for them is the same
(and unique), and it satisfies identity (\ref{HIROTA02}).
In this case the symmetry transformation connected with
operator $\hat R$ is correctly defined.
It is also possible to define one-parametric groups of
transformations by the equation
\bea
(\psi(\dots,\mu;g_1,\Theta_1)|\exp((\Theta_1-\Theta_2)\hat r)|
\psi(\lambda,\dots;g_2,\Theta_2))=0.
\label{H1}
\eea
Taking the generators $\hat r_{mn}=\lambda^n\partial_\lambda^m$,
we get noncommutative
symmetries in the form proposed by Orlov and Shulman \cite{Orlov}.
In our work we will consider non-isospectral symmetries connected with
operators with degenerate kernel, and, in particular,
generators with the kernel of the form
\be
r_{\alpha\beta}(\nu,\nu')={1\over 2\pi\text{i}} \delta(\alpha-\nu)
\delta(\beta-\nu'),
\label{delta}
\ee
where $\alpha,\beta$ belong to the unit circle, or, more generally,
\be
r_{\rho\wt\rho}(\nu,\nu')={1\over 2\pi\text{i}}\wt\rho(\nu')\rho(\nu),
\label{rho}
\ee
where for simplicity we put
$$
(\wt\rho|\rho)=0.
$$
{\bf Remark.} To make a transformation from the
generators  $\hat r_{\alpha\beta}$ to the generators
$\hat r_{mn}$ used by Orlov and Shulman, it is enough to note
that operator $\hat r_{\alpha\beta}$ can be represented
as a composition of the shift operator
$T_{\beta-\alpha}:\nu\rightarrow \nu+\beta-\alpha$
and operator of multiplication by the function
$\delta(\nu-\alpha)$. Then, expanding shift
operator and $\delta$-function into powers of parameters,
it is possible to make a transformation from one set
of generators to the other.

Using simple identity
$$
\exp(\Theta_{\alpha\beta} \hat r_{\alpha\beta})=
I+\Theta_{\alpha\beta} \hat r_{\alpha\beta},
$$
which is satisfied due to nilpotence of the generators,
and performing integration in the equation (\ref{H1}) taken for
$g_1=g_2$,
which in this case reads
\bea
\oint\oint d\nu d\nu'
\psi(\nu,\mu;g,\Theta_1)(\delta(\nu-\nu')+{\Theta_1-\Theta_2\over
2\pi\text{i}}\delta(\beta-\nu')\delta(\alpha-\nu))&&
\nn\\
\times
\psi(\lambda,\nu';g,\Theta_2))=0,&&
\label{H2}
\eea
we get equation for the CBA function
\bea
&&
\psi(\lambda,\mu;\bbox{x},\Theta_{\alpha\beta}+\Delta\Theta_{\alpha\beta})
=
\psi(\lambda,\mu;\bbox{x},\Theta_{\alpha\beta})
\nn\\
&&
\qquad
+\Delta\Theta_{\alpha\beta}
{\psi(\lambda,\beta;\bbox{x},\Theta_{\alpha\beta})
\psi(\alpha,\mu;\bbox{x},\Theta_{\alpha\beta}+\Delta
\Theta_{\alpha\beta})}.
\label{Delta}
\eea
It is possible to resolve this equation and express
$\psi(\lambda,\mu;\bbox{x},
\Theta_{\alpha\beta}+\Delta\Theta_{\alpha\beta})$ through
$\psi(\lambda,\mu;\bbox{x},\Theta_{\alpha\beta})$.
First we take equation (\ref{Delta}) at $\lambda=\alpha$
and get the expression for
$\psi(\alpha,\mu;\bbox{x},\Theta_{\alpha\beta}+\Delta\Theta_{\alpha\beta})$,
\bea
\psi(\alpha,\mu;\bbox{x},\Theta_{\alpha\beta}+\Delta\Theta_{\alpha\beta})=
{\psi(\alpha,\mu;
\bbox{x},\Theta_{\alpha\beta})\over 1-\Delta\Theta_{\alpha\beta}
\psi(\alpha,\beta;\bbox{x},\Theta_{\alpha\beta})}.
\label{Delta1}
\eea
Substituting (\ref{Delta1}) into (\ref{Delta}), we
finally get
\bea
&&
\psi(\lambda,\mu;\bbox{x},\Theta_{\alpha\beta}+\Delta\Theta_{\alpha\beta})=
\psi(\lambda,\mu;\bbox{x},\Theta_{\alpha\beta})
\nn\\&&\qquad\qquad
+\Delta\Theta_{\alpha\beta}
{\psi(\lambda,\beta;\bbox{x},
\Theta_{\alpha\beta})\psi(\alpha,\mu;\bbox{x},\Theta_{\alpha\beta}
)\over1-\Delta\Theta_{\alpha\beta}
\psi(\alpha,\beta;\bbox{x},\Theta_{\alpha\beta})}.
\label{Delta2}
\eea
In particular, this formula expresses
the function $\psi(\lambda,\mu;\bbox{x},\Theta_{\alpha\beta})$ through
the initial data $\psi_0(\lambda,\mu;\bbox{x})=
\psi(\lambda,\mu;\bbox{x},\Theta_{\alpha\beta}=0)$, thus giving explicit
formula for the action of non-isospectral symmetry connected
with the generator (\ref{delta}) on the CBA function.

Formula (\ref{Delta2}) can be rewritten as
\bea
&&
\psi(\lambda,\mu;\bbox{x},\Theta_{\alpha\beta})
\nn\\&&\qquad
=
\psi_0(\lambda,\mu;\bbox{x})
{1-\Theta_{\alpha\beta}{\text{det}_{\alpha\beta}\psi_0(\lambda,\mu;\bbox{x})
\over \psi_0(\lambda,\mu;\bbox{x})}
\over 1-\Theta_{\alpha\beta}\psi(\alpha,\beta;\bbox{x})},
\eea
where
\bea
\text{det}_{\alpha\beta}\psi_0(\lambda,\mu;\bbox{x})=
\text{det}\left(
\begin{array}{cc}
\psi_0(\lambda,\mu;\bbox{x})&\psi_0(\lambda,\beta;\bbox{x})\\
\psi_0(\alpha,\mu;\bbox{x})&\psi_0(\alpha,\beta;\bbox{x})
\end{array}
\right).
\eea
Recalling determinant formula for the transformation
of CBA function under the action of a rational loop
(see \cite{NLS}, \cite{AB1}),
\bea
\psi_0(\alpha,\beta;\bbox{x}+[\mu]-[\lambda])
={\text{det}_{\lambda\mu}\psi_0(\alpha,\beta;\bbox{x})
\over \psi_0(\lambda,\mu;\bbox{x})},
\eea
we get another representation of the transformation
(\ref{Delta2}),
\bea
\psi(\lambda,\mu;\bbox{x},\Theta_{\alpha\beta})=
\psi_0(\lambda,\mu;\bbox{x})
{1-\Theta_{\alpha\beta}\psi_0(\alpha,\beta;\bbox{x}+[\mu]-[\lambda])
\over 1-\Theta_{\alpha\beta}\psi_0(\alpha,\beta;\bbox{x})}.
\eea
Comparing this formula with the formula connecting the CBA
function and the $\tau$-function (which in fact defines
the $\tau$-function through the CBA function)
\bea
\psi(\lambda,\mu,\bbox{x})=
{g(\lambda)g(\mu)^{-1}}{1\over
\lambda-\mu}{\tau(\bbox{x}+[\mu]-[\lambda]) \over\tau(\bbox{x})},
\eea
we come to the conclusion that the $\tau$-function corresponding
to the transformed CBA function
$\psi(\lambda,\mu;\bbox{x},\Theta_{\alpha\beta})$
is given by the expression
\bea
\tau(\bbox{x},\Theta_{\alpha\beta})=
\tau_0(\bbox{x})(1-\Theta_{\alpha\beta}\psi_0(\alpha,\beta;\bbox{x})).
\label{tautrans}
\eea
Thus we have explicitly defined action of non-isospectral symmetry with
the generator (\ref{delta}) on KP $\tau$-function. Transformation
(\ref{tautrans}) coincides with the solitonic transformation
of the $\tau$-function defined through Date-Jimbo-Kashiwara-Miwa
vertex operator \cite{Date}.
Below we will demonstrate that in terms of potential
this is just a binary B\"acklund transformation, and for some
choice of KPSM solution $\Phi(\bbox{x})$ this is a M\"obius
transformation.

For the function $\Phi_{\alpha\beta}=\psi(\alpha,\beta;\bbox{x})$
satisfying
singular manifold equation (\ref{KPSM}) from  the formula (\ref{Delta2})
we get especially
simple transformation,
\bea
\Phi_{\alpha\beta}(\bbox{x},\Theta_{\alpha\beta})=
{\Phi^0_{\alpha\beta}(\bbox{x})\over1-
\Theta_{\alpha\beta}
\Phi^0_{\alpha\beta}(\bbox{x})},
\label{Delta3}
\eea
and this is nothing more then one-parametric subgroup
of the M\"obius group. Taking this formula
at $\Theta_{\alpha\beta}\rightarrow\infty$, we get (up to a constant)
transformation of inversion $\Phi_{\alpha\beta}\rightarrow
\Phi^{-1}_{\alpha\beta}$.

It is easy to check that the same
derivation holds for the generators (\ref{rho})
$\hat r_{\rho\tilde\rho}$, and in this case we get M\"obius
transformation
\bea
\Phi(\bbox{x},\Theta)=
{\Phi^0(\bbox{x})\over1-
\Theta
\Phi^0(\bbox{x})},
\label{Delta4}
\eea
for the solution of KPSM equation (\ref{KPSM})
corresponding to the weight functions $\rho(\nu)$,
$\wt\rho(\nu)$
\bea
\Phi(\bbox{x})=\oint\!\!\oint (\psi(\lambda,\mu;\bbox{x}))
\rho(\lambda)
\wt\rho(\mu) d\lambda\,d\mu.
\nn
\eea
\section{M\"obius Symmetry}
In this section we will concentrate on M\"obius symmetry of
KPSM equation (\ref{KPSM}), using only equations of generalized
hierarchy and connections between them, without explicit use of
bilinear technique underlying the construction. We will demonstrate
that M\"obius symmetry on the level of KPSM hierarchy generates
binary B\"acklund transformations on the level of the basic KP
hierarchy.

Characteristic feature of singular manifold equation
(\ref{KPSM}) is its invariance under M\"obius transformation
$$
\Phi\rightarrow {a\Phi+ b\over c\Phi+d},
$$
which can be easily checked. Now we are going to find the symmetry
of the basic KP hierarchy (\ref{KPa1}), which corresponds to the
M\"obius transformation on the level of the singular manifold
equation. To do that, we define the transformations of the
wave functions $\Psi$, $\wt \Psi$
using the equations (\ref{Phi1}),
and then we substitute the wave functions into linear equations
(\ref{KPbasea13}) and (\ref{KPbaseb13})
to find the transformation of the potential $u$. Generic
M\"obius transformation can be represented as composition of
translation, scaling and inversion. Translation and scaling of $\Phi$
do not change the potential $u$ (translation doesn't change wave
functions, and scaling of wave functions doesn't change the
potential), and so in principle our problem is to find the
transformation of potential $u$ corresponding to inversion
$
\Phi\rightarrow {\Phi}^{-1}
$.
Transformations of the wave functions, according to equations
(\ref{Phi1}), look like
$
\wt\Psi\rightarrow -\Phi^{-1}\wt \Psi$, $\Psi\rightarrow
\Phi^{-1}\Psi
$,
and, substituting them into linear equations (\ref{KPbasea13})
and (\ref{KPbaseb13}),
we get the formula for the transformation of
potential $u$,
\be
u(\Phi^{-1})=u(\Phi)- \Psi\Phi^{-1}\wt\Psi.
\label{utrans}
\ee
Taking into account equation
\be
\partial_x\Phi=\wt\Psi\Psi,
\label{derivative}
\ee
arising in the zeroth order of expansion of equation (\ref{Phi1}),
it is possible to rewrite formula (\ref{utrans}) as
\be
u({\Phi}^{-1})=u(\Phi)-\partial_x\ln \Phi,
\label{utrans1}
\ee
which is a well-known binary B\"acklund transformation.
Thus we have shown that inversion on the level
of KPSM equation hierarchy leads to binary B\"acklund transformation
on the level of the basic KP hierarchy. The connection
between M\"obius transformation and binary B\"acklund transformation
was discovered in the framework of Painleve analysis \cite{Weiss}.

Let us consider a one-parametric subgroup  of the M\"obius group
\be
\Phi(\Theta)={\Phi_0\over 1-\Theta \Phi_0}
\label{Phicont}
\ee
characterized by the equation
\be
\partial_\Theta \Phi=\Phi^2.
\label{Phicont1}
\ee
Using equation (\ref{utrans1}), it is easy to find continuous
symmetry of KP hierarchy corresponding to this subgroup.
First, directly from  (\ref{utrans1}) we get a formula
\be
u(\Theta):=u\left({\Phi_0\over 1-\Theta \Phi_0}\right)=
u\left({1-\Theta \Phi_0\over\Phi_0}\right)-\partial_x\ln
\left({1-\Theta \Phi_0\over\Phi_0}\right).
\label{utrans11}
\ee
Taking into account that
$$
u\left({1-\Theta \Phi_0\over\Phi_0}\right)=u(\Phi_0^{-1}),
$$
and transforming $u(\Phi_0^{-1})$ using formula
(\ref{utrans1}), we finally find symmetry transformation
of potential $u$ depending on continuous parameter $\Theta$,
\be
u(\Theta)=u_0-\partial_x\ln(1-\Theta\Phi_0).
\label{trans}
\ee
Potential $u$ satisfies differential relation
\be
\partial_\Theta u=\partial_x\Phi,
\label{difftrans}
\ee
or, taking into account formula (\ref{derivative}),
\be
\partial_\Theta u=\Psi\wt\Psi.
\label{difftrans1}
\ee
Expression $\Psi\wt\Psi$ represents infinitesimal
(in general non-isospectral) symmetry of KP hierarchy \cite{Orlov1},
and the formula (\ref{trans}) defines a one-parametric group of
transformations connected with this symmetry (specified by
extra relation (\ref{Phicont1})). Exactly this form of
symmetry generator is used to define constrained KP hierarchy
\cite{KS0}, \cite{KS},
which will be one of the objects of our study.

We will also consider more general symmetry transformations
of KPSM hierarchy (\ref{KPSM}), which we call multicomponent
M\"obius-type transformations. We have used an arbitrary pair of wave
functions $\Psi$, $\wt\Psi$ to define the function $\Phi$ through the set
of equations (\ref{Phi1}). Let us fix a set of wave functions and dual
wave functions $\Psi_k$, $\wt\Psi_k$, $1\leq k\leq \infty$
(in terms of Hirota bilinear identity
we should fix a set of weight functions $\rho_k(\nu)$,
$\wt\rho_k(\nu)$).
Then equations (\ref{Phi1}) define a matrix of solutions
of equation (\ref{KPSM}) $|\Phi|$ connected with the same solution
$u$ of KP hierarchy (\ref{KPa1}).
Matrix
entries $\Phi_{kp}$ satisfy the equations
\be
{\Delta_i\over a_i}\Phi_{kp}=\wt\Psi_k T_i \Psi_p,
\ee
or, in matrix form,
\be
{\Delta_i\over a_i}|\Phi|=|\wt\Psi\rangle T_i \langle\Psi|.
\label{transmulti}
\ee
It is easy to check that matrix inversion
$
|\Phi|\rightarrow|\Phi|^{-1}
$
leads to the same equation (\ref{transmulti})
with transformed $|\wt\Psi\rangle$, $\langle\Psi|$,
$
|\wt\Psi\rangle\rightarrow |\Phi|^{-1}|\wt\Psi\rangle$,
$\langle\Psi|\rightarrow \langle\Psi||\Phi|^{-1}$.
Substituting transformed vectors of wave functions
into linear equations (\ref{KPbasea13}), (\ref{KPbaseb13}),
we come to the conclusion that all components of the wave functions
give the same transformed potential
\be
u\rightarrow u -\langle\Psi|\Phi^{-1}|\wt\Psi\rangle.
\label{transmulti1}
\ee
Taking into account that equation
$
\partial_x|\Phi|=|\wt\Psi\rangle\langle\Psi|
$
imply the identity
$$
\langle\Psi|\Phi^{-1}|\wt\Psi\rangle=\partial_x\ln\det |\Phi|,
$$
we get another form of transformation of potential
corresponding to multicomponent M\"obius-type transformation,
\be
u(|\Phi|^{-1})=u(|\Phi|)-\partial_x\ln\det |\Phi|,
\label{transmulti2}
\ee
which represents a composition formula for several binary B\"acklund
transformations.

Multicomponent continuous M\"obius-type symmetry
$$
|\Phi(\Theta)|=|\Phi_0|(I-\Theta|\Phi_0|)^{-1},\qquad
\partial_\Theta |\Phi|=|\Phi||\Phi|,
$$
leads to continuous symmetry for the potential
\bea
u(\Theta)=u_0-\partial_x\ln\det(I-\Theta|\Phi_0|),
\label{transmulti3}
\\
\partial_\Theta u=\partial_x \text{tr}|\Phi|=\sum_{i=1}^N \Psi_i\wt\Psi_i.
\label{diff2}
\eea
\section{Symmetry Constraints and Calogero-Moser System}
The concept of generalized hierarchy is rather effective tool
in the study
of symmetry constraints.
A standard symmetry constraint for KP hierarchy is
\cite{KS0}, \cite{KS}
\be
u_x=\Psi\wt\Psi,
\label{constraint}
\ee
and it was shown in \cite{KS0} that it leads
to AKNS hierarchy for the wave functions. It is possible also to derive
two-dimensional equation for one function (either $u$ or $\Phi$).
Indeed, it was shown above that
$
\Phi_x=\Psi\wt\Psi
$,
so for the constrained hierarchy
$$
u_x=\Phi_x.
$$
Thus $u$ and $\Phi$ represent almost the same object, and the meaning
of the constraint is that it glues the first and the third level of
generalized hierarchy. We know that $u$ satisfies KP equation
(\ref{KP00}), and
$\Phi$ satisfies KPSM equation, but, using the constraint, we can
write two equations for both of these functions. Combining these
equations, it is easy to eliminate the terms containing partial
derivative over $t$ and get two-dimensional differential relation,
for $\Phi$
it looks like (see also \cite{KS0})
\bea
\partial_x\left(
\partial_x\bigl(\mbox{${3\over2}$}(\Phi_x)^2-
\mbox{${3\over8}$} {\frac{
\Phi_y^2-\Phi_{xx}^2}{\Phi_x}}\bigr)+
\mbox{${3\over4}$}{\Phi_y\Phi_{xy}\over\Phi_x}-
\mbox{${3\over4}$}\Phi_{xx}W_y\right)=0,
\quad W_x={
\frac{\Phi_y}{\Phi_x}}.
\nn
\eea

Let us consider one-parametric group of symmetry transformations
of potential $u$ defined by the formula (\ref{transmulti3});
we have shown that $u$ satisfies differential relations
\bea
\partial_\Theta u=\partial_x \text{tr}|\Phi|=\sum_{i=1}^N \Psi_i\wt\Psi_i.
\nn
\eea
According to these relations, standard constrains of the type
(\ref{constraint}) can be interpreted as an equation
\be
u_\Theta=u_x,\quad \Theta=0.
\label{constraint2}
\ee

There is stronger symmetry constraint, for which
relation (\ref{constraint2}) is required
to be satisfied {\em for all}
$\Theta$, not only at the origin. The dependence of $u$ on extra
time $\Theta$ is rational, so the constraints of this type impose
rational dependence of $u$ on $x$. In this way we come to
rational Calogero-Moser system. Indeed, let us make a simple
transformation of the
formula (\ref{transmulti3}) using relation (\ref{transmulti2}),
\bea
u(\Theta)=u(|\Phi_0|^{-1})-\partial_x\ln\det(|\Phi_0|^{-1}-\Theta),
\label{transmulti4}
\eea
and substitute the result to the equation (\ref{constraint2}).
Comparing the singularities, we come to the conclusion that
\be
v=-2u_x=\sum_{i=1}^N{-2\over(\phi_i(y,\cdots)-x-\Theta)^2},
\label{Calogero}
\ee
where $\phi_i$ are eigenvalues of the matrix
$|\Phi(x=0,y,t,\cdots)|^{-1}$. Due to the constraint the eigenvalues of
this matrix should depend linearly on $x$,
$
\phi_i(x)=\phi_i(0)-x
$, and also $\partial_x u(|\Phi_0|^{-1})=0$.
The substitution for the potential  (\ref{Calogero})
characterizes Calogero-Moser integrable system of particles
on the line \cite{Krichever}. Thus we have demonstrated that this system can
be obtained through the symmetry constraint of KP hierarchy.
\section*{Acknowledgments}
The first author (LB) is grateful to the
Dipartimento di Fisica dell' Universit\`a
and Sezione INFN, Lecce, for hospitality and support;
(LB) also acknowledges partial support from the
Russian Foundation for Basic Research under grants
No 98-01-00525 and 96-15-96093 (scientific schools).


\begin{thebibliography}{99}
\bibitem{AB1}
L.V. Bogdanov  and B.G. Konopelchenko,
\Journal{\JMP}{39(9)}{4683}{1998}.
\bibitem{AB2}
L.V. Bogdanov and B.G. Konopelchenko,
\Journal{\JMP}{39(9)}{4701}{1998}.
\bibitem{dbar}
V.E. Zakharov and S.V. Manakov,
\Journal{\LOMI}
{133}{77}{1984} [in Russian].
\\
V.E. Zakharov and S.V. Manakov,
\Journal{\FAP}{19(2)}{11}{1985}
[in Russian].
\bibitem{NLS}
L.V. Bogdanov, \Journal{\PHD}{87(1-4)}{58}{1995}.

\bibitem{Sato}
M. Sato,
\Journal{RIMS, Kokyuroku, Kyoto Univ.}{439}{30}{1981},
\\
\Proc{M. Sato and Y. Sato}{}
{Nonlinear partial differential equations in applied science,
eds. {H. Fujita et al}}
{259}
{North-Holland, Amsterdam-New York}{1983}.
\bibitem{Wilson}
G. Segal and G. Wilson,
\Journal{Inst. Hautes \'Etudes Sci. Publ. Math.}
{61}{5}{1985},\\
\Book{G. Segal and A. Pressley}{Loop groups}
{Clarendon, Oxford}{1986}.
\bibitem{Orlov}
A.Yu. Orlov and E.I. Shulman,
\Journal{\LMP}{12}{171}{1986}.
\bibitem{Date}
\Proc{E. Date, M. Kashiwara, M. Jimbo, T. Miwa}{}
{Nonlinear integrable systems---classical theory and quantum theory,
eds. {M. Jimbo and T. Miwa}}
{39}{World Scientific Publishing,
Singapore}
{1983}.
\bibitem{Weiss}
J. Weiss, \Journal{\JMP}{24}{1405}{1983}.

\bibitem{Moeb}
L.V. Bogdanov and B.G. Konopelchenko,  \Journal{\PLA}{256}{39}{1999}.
\bibitem{KS0}
B.Konopelchenko and W.Strampp,
\Journal{Inverse Problems}{7}{L17}{1991}.
\bibitem{KS}
B.Konopelchenko and W.Strampp,
\Journal{\JMP}{33}{3676}{1992}.
\bibitem{Orlov1}
\Proc{A. Yu. Orlov}{}
{Plasma Theory and Nonlinear and Turbulent Processes in Physics,
ed. {Baryakhtar}}
{116}{World Scientific Publishing,
Singapore}
{1988}.

\bibitem{Krichever} I.M. Krichever,
\Journal{Funct. Anal. Appl.}{12}{59}{1978}.
\end{thebibliography}
\end{document}